\documentclass{aa}  

\usepackage{txfonts}
\usepackage{hyperref}
\usepackage{graphicx}	
\usepackage{amsmath}	
\usepackage{amssymb}	
\usepackage{natbib}
\usepackage{longtable}
\usepackage{float}
\usepackage{threeparttable}
\usepackage{subcaption}
\usepackage{booktabs,caption}
\usepackage{soul}
\usepackage[table]{xcolor}
\newcommand{\hours}{^\text{h}}
\newcommand{\minutes}{^\text{m}}
\newcommand{\seconds}{^\text{s}}
\include{def_bibtex}

\begin{document} 

 \title{Quiescent X-ray variability in the neutron star Be/X-ray transient GRO~J1750-27}

   \author{A.~Rouco~Escorial
          \inst{1}\fnmsep\thanks{A.RoucoEscorial@uva.nl}
          \and
           R.~Wijnands\inst{1}
          \and
          L.~S.~Ootes\inst{1}
          \and
          N.~Degenaar\inst{1}
          \and
          M.~Snelders\inst{1}
          \and
          L.~Kaper\inst{1}
          \and
          E.~M.~Cackett\inst{2}
          \and
          J.~Homan\inst{3,4}
          }
          
   \institute{Anton Pannekoek Institute for Astronomy, University of Amsterdam, Science Park 904, 1098 XH, Amsterdam, The Netherlands
         \and
             Department of Physics and Astronomy, Wayne State University, 666 W. Hancock St, Detroit, MI 48201, USA
         \and
         	Eureka Scientific, Inc., 2452 Delmer Street, Oakland, CA 94602, USA
         \and
         	SRON, Netherlands Institute for Space Research, Sorbonnelaan 2, 3584 CS Utrecht, The Netherlands
             }
             
   \date{Received; accepted}
%

\abstract
  {The Be/X-ray transient GRO~J1750-27 exhibited a type-II (giant) outburst in 2015. After the source transited to quiescence, we triggered our multi-year \textit{Chandra} monitoring programme to study its quiescent behaviour. The programme was designed to follow the cooling of a potentially heated neutron-star crust due to accretion of matter during the preceding outburst, similar to what we potentially have observed before in two other Be/X-ray transients, namely 4U~0115+63 and V~0332+53. However, unlike for these other two systems, we do not find any strong evidence that the neutron-star crust in GRO~J1750-27 was indeed heated during the accretion phase. We detected the source at a rather low X-ray luminosity ($\sim$10$^{33}$\,erg~s$^{-1}$) during only three of our five observations. When the source was not detected it had very low-luminosity upper limits  (<\,$10^{32}$\,erg~s$^{-1}$; depending on assumed spectral model). We interpret these detections and the variability observed as emission likely due to very low-level accretion onto the neutron star. We also discuss why the neutron-star crust in GRO~J1750-27 might not have been heated while the ones in 4U~0115+63 and V~0332+53 possibly were.}
  
%

\titlerunning{The variable X-ray behaviour of GRO~J1750-27 in quiescence}
\authorrunning{A. Rouco Escorial et al.}
\maketitle
%

\section{Introduction}\label{sec:P3_GROJ1750_introduction}
 
Be/X-ray binary systems are the most common sub-type of high-mass X-ray binaries in which magnetised neutron stars (NSs; with a magnetic field of B$\sim$10$^{12-13}$\,G) accrete from their massive companions (a Be-type star in our case) while moving around them in (highly) eccentric orbits. These Be/X-ray binaries show two kinds of transient X-ray behaviour (for a review of these systems see \citealt{Reig2011}): type-I (normal) and type-II (giant) outbursts. The type-I outbursts, which have a short duration (a fraction of an orbital period), are caused by the accretion of matter onto the NS when the compact object passes through the decretion disk of the companion during the periastron passage. The X-ray luminosity (L$_\textnormal{X}$) related to these events usually peaks at L$_\textnormal{X}$$\sim$10$^{36-37}$\,erg~s$^{-1}$. On the contrary, the type-II outbursts normally (although not always) last for more than an orbital period and are very bright, reaching or even exceeding the Eddington limit for a NS (L$_\textnormal{X}$\,>\,2$\times10^{38}$\,erg~s$^{-1}$). The physical mechanism behind these giant, type-II outbursts remains unclear, although several studies have approached the problem by focusing on the structure of the Be-star decretion disk and its alignment with the NS orbit (\citealt{Moritani2013}; \citealt{Martin2014}; \citealt{Monageng2017}) or by studying the effects of perturbations in the decretion disk \citep{Laplace2017}.

The bright active episodes of Be/X-ray transients are powered by the accretion of matter onto the NS. If the accretion rate is high, the matter can overcome the magnetospheric barrier of the NS and the material is channeled toward the magnetic poles. At the end of the outbursts, when the mass accretion rate decreases, the NS spin becomes a decisive component in the accretion process. In the case of relatively fast spinning systems (with typical spin periods, P$_\textnormal{spin}$\,<\,10\,--\,100\,s; depending on the exact strength of the surface magnetic field of the NS) the ram pressure of the matter in the accretion flow is unable to overcome the magnetospheric barrier. It is generally thought that this material is then expelled from the inner part of these systems through what is called the `propeller effect' (\citealt{Illarionov1975}; \citealt{Romanova2004}; \citealt{DAngelo2010}). If the propeller effect is not very strong, the matter might also accumulate outside of the magnetosphere in what is called a `trapped' or 'dead disk' (see, e.g., \citealt{Syunyaev1977}; \citealt{Dangelo2012}; \citealt{Patruno2013}; \citealt{Dangelo2014}).

In the case of the very fast spinning systems (P$_\textnormal{spin}$\,<\,10\,s), the expected luminosity below which they are assumed to be in the propeller regime is L$_{\textnormal{X}_\textnormal{prop}}$$\sim$10$^{35-36}$\,erg~s$^{-1}$. Since in this regime no matter is thought to accrete anymore on the NS, these systems are expected to be very dim in quiescence. Indeed, these systems have quiescent luminosities of only $\sim$10$^{32-33}$\,erg~s$^{-1}$. However, it is very likely that this low-level emission of X-rays does not have a single origin. In some systems, there is strong evidence that, despite being in the propeller regime, low-level accretion onto their NS surfaces still continues. This indicates that the propeller effect might not always be completely effective although how matter exactly reaches the NS surface is still unclear (see discussions in \citealt{Orlandini2004}; \citealt{Mukherjee2005}; \citealt{Doroshenko2014}). Another possible mechanism that could produce low-level emission in the propeller regime is the accretion flow at the magnetospheric boundary. This flow could produce significant radiation and might be detectable at luminosities of $\sim$10$^{32-34}$\,erg~s$^{-1}$ \citep{Campana2001a}. However, it is unclear exactly how the emission would be generated (see \citealt{Ikhsanov2001} and \citealt{Lii2014} for discussion) and likely most of the released energy will not be emitted in the X-rays but at longer wavelengths such as the ultraviolet (see discussion in \citealt{Tsygankov2016}).

If no matter reaches the NS surface when Be/X-ray transients are in the propeller regime, it might be possible that the NS becomes visible at L$_\textnormal{X}$$\sim$10$^{32-34}$\,erg~s$^{-1}$ due to thermal emission from its surface. During the outburst, matter is deposited on the NS surface and compresses the inner layers of the crust, triggering nuclear reactions that release heat deep in the crust (e.g., \citealt{Haensel1990, Haensel2003, Haensel2008}; \citealt{Steiner2012}; \citealt{Lau2018}). This release of energy heats up the crust, which can become out of thermal equilibrium with the NS core if enough energy is generated during an outburst (e.g., \citealt{Rutledge2002}). Once the outburst is over and the accretion has halted, the heat is conducted both inwards to the core and outwards where it is emitted as cooling emission from the surface untill the crust-core thermal equilibrium is restored again. This process has been observed in about a dozen of accreting low-magnetic field NSs in low-mass X-ray binaries (see review of \citealt{Wijnands2017}) and potentially could also be observed for high-magnetic field accreting NSs. Indeed, evidence for this process has been observed in two such systems (4U~0115+63 and V~0332+53; e.g., \citealt{Wijnands2016}; \citealt{Rouco2017a}), although it still needs to be confirmed if indeed the cooling of the accretion-heated NS crusts was the dominant emission process. The cooling-time scale of the crust in such high-magnetic field NS systems is unclear (and might be relatively short; see discussion in \citealt{Rouco2017a} and \citealt{Tsygankov2017b}), but when it is again in equilibrium with the core, thermal emission from the surface might be still observable if the NS core is hot enough. Such surface emission has been inferred for several systems (see, e.g., \citealt{Campana2002}; \citealt{Reig2014}; \citealt{Elshamouty2016}; \citealt{Tsygankov2017b}).

In order to investigate further the emission processes potentially at work in Be/X-ray transients when not in outburst,  we study the behaviour of GRO~J1750-27 (also known as AX~J1749.1-2639) after its 2015 type-II outburst. GRO~J1750-27 is a Be/X-ray transient that harbours a 4.45\,s pulsar (\citealt{Bildsten1997}), which orbits around its companion star every 29.8 days (\citealt{Scott1997}). The orbit has an eccentricity of $\sim$0.3. The source was discovered by the Burst And Transient Source Experiment on board of the \textit{Compton Gamma Ray Observatory} in 1997 \citep{Scott1997} and studied further by \citet{Shaw2009} during its second observed outburst in 2008. It remained dormant until a new type-II outburst started at the beginning of 2015 (\citealt{Finger2014}).

\section{Observations, analysis and results}\label{sec:P3_GROJ1750_analysis}

\subsection{Observations and data reduction}\label{subsec:P3_GROJ1750_observations}

%
\begin{figure*}[h]
	\centering
	\includegraphics[width=2.0\columnwidth]{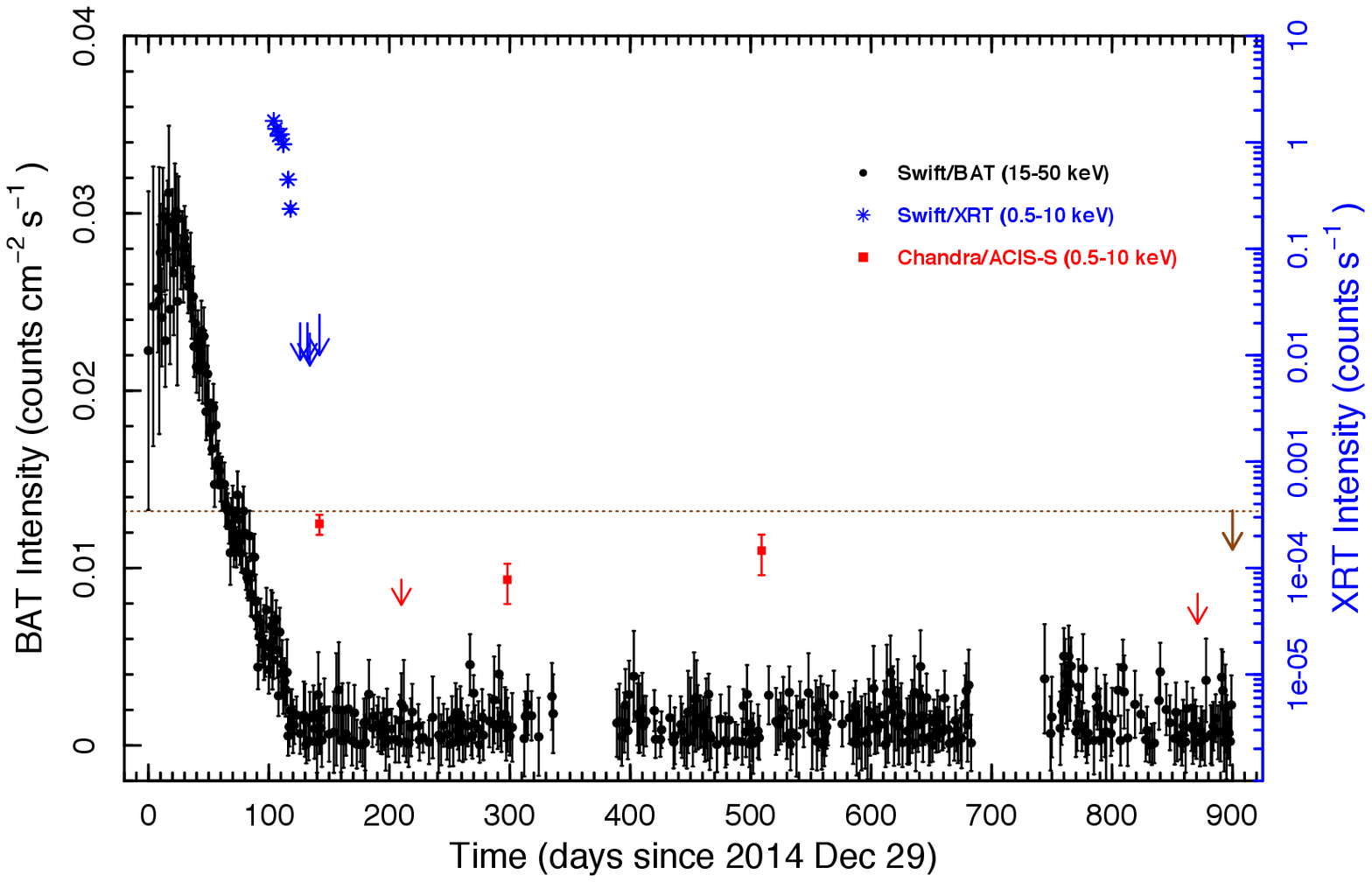}
    \caption{The \textit{Swift}/XRT (blue) and \textit{Swift}/BAT (black) light curves during and after the 2015 type-II outburst of GRO~J1750-27. All \textit{Chandra} count rates have been converted to \textit{Swift}/XRT count rates using the method described in Sec.\,\ref{subsec:P3_GROJ1750_observations}. The \textit{Chandra} detections (with 1$\sigma$ errors) are given as red squares. The \textit{Chandra} upper limits are determined in the manner explained in Sec.\,\ref{subsec:P3_GROJ1750_observations} and shown as red arrows. The dotted brown line and brown arrow represent the upper limit obtained in the \textit{Chandra} observation 14643 taken in May 2013, i.e., almost a year and a half before the 2015 type-II outburst. The zeropoint of the light curve (the fiducial starting point of the outburst) corresponds to December 29th, 2014.}
    \label{fig:P3_GROJ1750_BATlightcurve}
\end{figure*}
%

%
\begin{table*}[h]
    \caption{Log of \textit{Swift} and \textit{Chandra} observations for GRO~J1750-27}
    \centering
    \begin{tabular}{llcccc}
        \hline\hline
        \noalign{\smallskip}
        \multicolumn{6}{c}{\textit{Swift}}  \\
        \noalign{\smallskip}
        \hline
        \noalign{\smallskip}
        ObsID   &   MJD &   Calendar Date   &   Exposure Time   &   Count Rate  &   \\
            &       &   (2015)  &   (ks)    &   (10$^{-2}$\,counts~s$^{-1}$) &   \\
        \noalign{\smallskip}
        \hline
        \noalign{\smallskip}
        000311150[12]   &   571[24] &   12-04   &   $\sim$0.8   &   159.1$\pm$8.6   &   \\
        +13 &   +26 &   14-04   &   $\sim$0.9   &   134.2$\pm$7.5   &   \\
    	+14 &   +28 &   16-04   &   $\sim$0.3   &   115$\pm$11  &   \\
    	+15 &   +30 &   18-04   &   $\sim$0.9   &   119.3$\pm$6.2   &   \\
    	+16 &   +32 &   20-04   &   $\sim$1.0   &   95.9$\pm$6.1    &   \\
    	+17 &   +36 &   24-04   &   $\sim$0.9   &   44.7$\pm$2.8    &   \\
    	+18 &   +38 &   26-04   &   "   &   23.7$\pm$1.8    &   \\
    	+20 &   +46 &   04-05   &   $\sim$1.0   &   <\,2.0  &   \\
    	+22 &   +52 &   10-05   &   $\sim$0.9   &   <\,2.0  &   \\
    	+23 &   +54 &   12-05   &   $\sim$1.0   &   <\,1.6  &   \\
    	+24 &   +62 &   20-05   &   $\sim$0.6   &   <\,2.4  &   \\
    	\noalign{\smallskip}
        \hline\hline
        \noalign{\smallskip}
        \multicolumn{6}{c}{\textit{Chandra}}    \\
        \noalign{\smallskip}
        \hline
        \noalign{\smallskip}
        ObsID   &   MJD &   Calendar Date   &   Exposure Time   &   Count Rate  &   $\phi$  \\
            &       &       &   (ks)    &   (10$^{-4}$\,counts~s$^{-1}$) &   \\
        \noalign{\smallskip}
        \hline
        \noalign{\smallskip}
		14643   &   56434   &   22-05-2013  &   $\sim$5 &   <\,12   &   0.22\\	
		167[23] &   57[162] &   20-05-2015  &   $\sim$27    &   9.0$\pm$1.9 &   0.65\\
		+24 &   +230    &   27-07-2015  &   "   &   <\,2.7  &   0.93\\
		+25 &   +318    &   23-10-2015  &   $\sim$29    &   2.8$\pm$1.2  &   0.89\\
		+26 &   +529    &   21-05-2016  &   "   &   5.1$\pm$1.5 &   0.97\\
		+27 &   +891    &   18-05-2017  &   "   &   <\,2.0  &   0.11\\
		\noalign{\smallskip}
		\hline
		\noalign{\smallskip}
		+[24,27] &   --  &   --  &   $\sim$55    &   <\,1.4  &   --\\
		\noalign{\smallskip}
		\hline\hline
    \end{tabular}
    \tablefoot{The \textit{Swift}/XRT count rates (with 1$\sigma$ errors) are for the 0.5-10\,keV energy range and XRT upper limits are calculated as mentioned in Sect.\,\ref{subsec:P3_GROJ1750_observations}. The \textit{Chandra} (background subtracted) count rates (with 1$\sigma$ errors) are given for the 0.5-7\,keV energy range and the upper limits are calculated as mentioned in Sect.\,\ref{subsec:P3_GROJ1750_observations} as well. $\phi$ represents the orbital phase of the binary when the observations were taken (with $\phi$=0 defined as periastron passage and $\phi$=0.5 as apastron). The last row of the table correspond to the count-rate upper limit in the combined observation.}
    \label{tab:P3_GROJ1750_log_observations}
\end{table*}
%

The Neil Gehrels \textit{Swift} observatory (from now on referred to as \textit{Swift}) monitored GRO~J1750-27 during its giant, type-II outburst in early 2015 using the Burst Alert Telescope (BAT) and the X-ray Telescope (XRT; see Fig.\,\ref{fig:P3_GROJ1750_BATlightcurve}). The \textit{Swift}/BAT data were obtained from the \textit{Swift}/BAT hard X-ray transient monitor web page\footnote{\url{https://swift.gsfc.nasa.gov/results/transients/weak/AXJ1749.1-2639/}} (\citealt{Krimm2013}) and the \textit{Swift}/XRT light curve from the \textit{Swift}/XRT products web interface\footnote{\url{http://www.swift.ac.uk/user_objects/}} (\citealt{Evans2009}) using our updated source position (see below). Unfortunately, the BAT missed the beginning of the type-II outburst and when it started to monitor the system the quality of the data was poor (see the large error bars on the BAT points in Fig.\,\ref{fig:P3_GROJ1750_BATlightcurve}). Therefore, the exact starting date of the outburst could not be obtained from the BAT and we used the date of the first good BAT point as fiducial starting point in Fig.\,\ref{fig:P3_GROJ1750_BATlightcurve}. The decay of the outburst was followed using the XRT (see Fig.\,\ref{fig:P3_GROJ1750_BATlightcurve} and Table\,\ref{tab:P3_GROJ1750_log_observations} for a log of the observations), but the instrument was not sensitive enough to detect the source during the final phases of the decay and its subsequent transition into quiescence (see Fig.\,\ref{fig:P3_GROJ1750_BATlightcurve}).

Once the source was no longer detected using the \textit{Swift}/XRT, our multi-year \textit{Chandra} monitoring campaign (PI: Wijnands) was triggered to investigate if cooling of a potential accretion-heated crust could be observed in this source. Our \textit{Chandra} campaign consisted of five observations that were performed between 2015 May 20 and 2017 May 18 (see Table\,\ref{tab:P3_GROJ1750_log_observations}). In addition, we report on a previous \textit{Chandra} observation with observation identification (ObsID) 14643 that was obtained on 2013 May 22. All our \textit{Chandra} observations were performed using the ACIS-S detector using the faint and timed detector mode. Typically a 1/4 subarray was used to limit the pile-up in case the source was unexpectedly bright, except in observation 16724 during which a 1/8 subarray was used.

We reduced and analysed the data using the CIAO tools (v.\,4.9)\footnote{\url{http://cxc.harvard.edu/ciao/}} and the CALDB (v.\,4.7.6)\footnote{\url{http://cxc.harvard.edu/caldb/}}. We reprocessed the data files following the standard procedures\footnote{\url{http://cxc.harvard.edu/ciao/guides/}} and inspected each observation for any possible background flares\footnote{\url{http://cxc.harvard.edu/ciao/threads/flare/}}. We did not find any period of high background, therefore all the data were used. We detected GRO~J1750-27 in three of our five \textit{Chandra} observations (ObsIDs 16723, 16725 and 16726; see Table\,\ref{tab:P3_GROJ1750_log_observations}) at a position of RA~(J2000)$=17\hours49\minutes12.96\seconds$ and Dec~(J2000)$=-26^{o}38^{\prime}38\,.\!\!^{\prime\prime}$6, with a 90$\%$ uncertainty radius of 0\,.$\!\!^{\prime\prime}$9. This position was obtained using the CIAO routine WAVDETECT with default parameter values (detection threshold of sigthresh=10$^{-6}$). Our \textit{Chandra} position falls well within the {\it Swift}/XRT error circle reported by \citet[][see left panel in our Fig.\,\ref{fig:P3_GROJ1750_image}]{Shaw2009} obtained when the source was in outburst, demonstrating that we conclusively have detected GRO~J1750-27 in quiescence.

When the source was detected, we used the same source and background extraction regions as we used for our spectral analysis (see Sect.\,\ref{subsec:P3_GROJ1750_spectra} for the details) to extract the count rates. In order to compare our {\it Chandra} count rates with the obtained {\it Swift}/XRT ones, we converted the {\it Chandra} count rates to so-called `inferred XRT count rates' (in the energy range 0.5-10\,keV) using the WEBPIMMS\footnote{\url{http://heasarc.gsfc.nasa.gov/cgi-bin/Tools/w3pimms/w3pimms.pl}} tool and the spectral parameters obtained from the first \textit{Chandra} detection in observation 16723 (see Sect.\,\ref{subsec:P3_GROJ1750_spectra}). When we did not detect the source, we calculated the 2$\sigma$ count-rate upper limits following the method described by \citet{Gehrels1986}. The obtained {\it Chandra} upper limits were converted to XRT upper limits following the steps mentioned previously.

\subsection{Light curve}\label{subsec:P3_GROJ1750_lightcurve}

%
\begin{figure*}
	\centering
	\includegraphics[width=2.0\columnwidth]{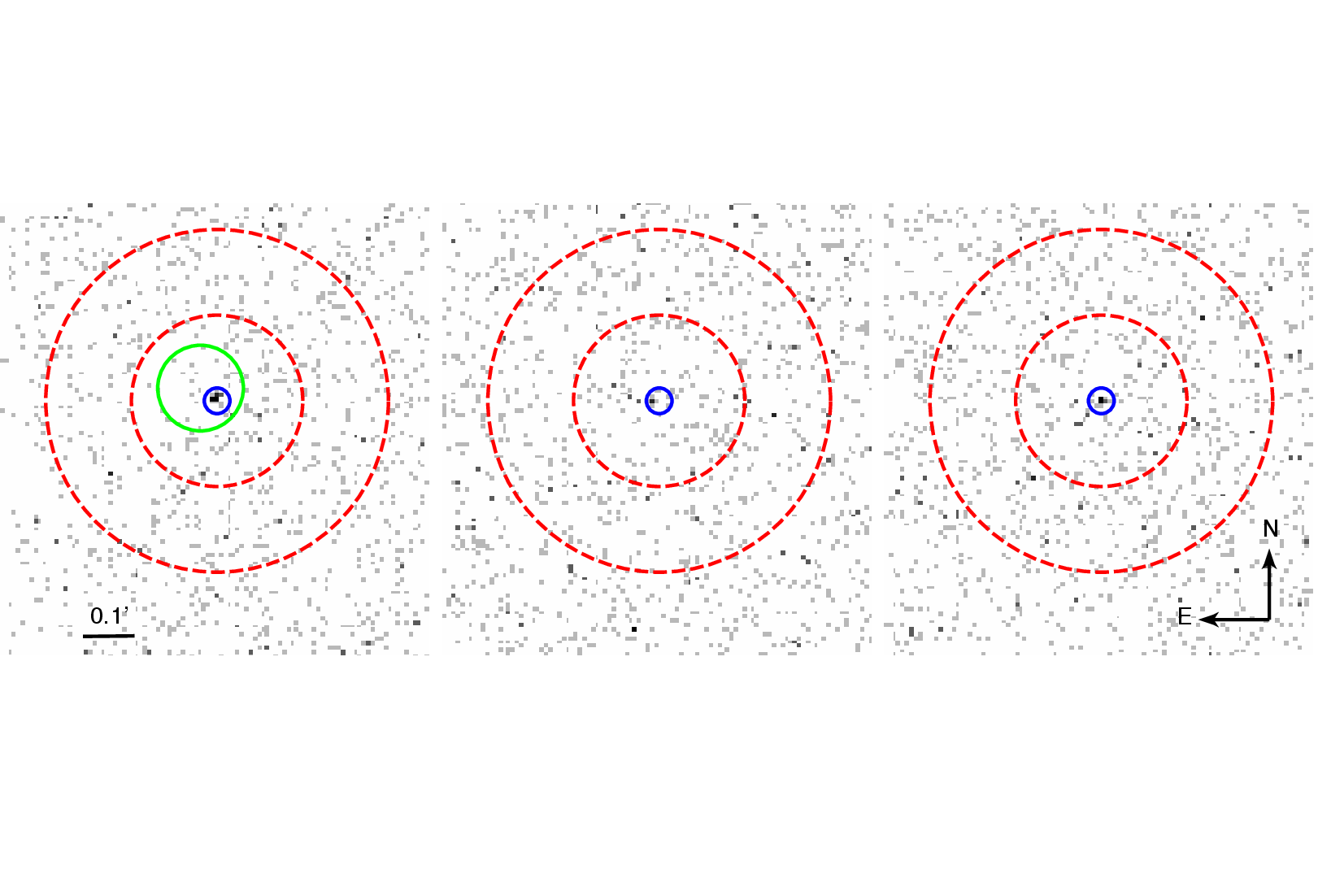}
    \caption{Close-up images (0.5-7\,keV energy range) of our \textit{Chandra} detections (listed by their ObsIDs; from left to right): 16723, 16725 and 16726. The blue circles indicate the source extraction regions and the dashed red annuli the background extraction regions. The green circle is the \textit{Swift}/XRT error region of the source reported by \citet{Shaw2009}. See Sect.\,\ref{subsec:P3_GROJ1750_observations} and Sect.\,\ref{subsec:P3_GROJ1750_lightcurve} for more details about the detections and upper limits.}
    \label{fig:P3_GROJ1750_image}
\end{figure*}
%

As we can see from Fig.\,\ref{fig:P3_GROJ1750_BATlightcurve}, the XRT count rate decreased from $\sim$1.6\,counts~s$^{-1}$ at the start of the XRT observations to <\,0.02\,counts~s$^{-1}$ during the last part of the XRT monitoring. At that time, our \textit{Chandra} programme had already been triggered and the first \textit{Chandra} observation was taken the same day as our last \textit{Swift} observation (see Table\,\ref{tab:P3_GROJ1750_log_observations} for a log of our \textit{Swift} and \textit{Chandra} observations). Due to the better sensitivity of \textit{Chandra} and the longer exposure time, the source was detected (see Fig.\,\ref{fig:P3_GROJ1750_image}, left panel) in this first {\it Chandra} observation (ObsID 16723) with a net count rate of (9.0$\pm$1.9)$\times10^{-4}$\,counts~s$^{-1}$ (24.3$\pm$5.2 net source photons; for the 0.5-7\,keV energy range) which resulted in an inferred XRT count rate of $\sim$2.6$\times$10$^{-4}$\,counts~s$^{-1}$. Thus the source was almost 3 orders of magnitude fainter than when it was last detected using the XRT (see Fig.\,\ref{fig:P3_GROJ1750_BATlightcurve} and Table\,\ref{tab:P3_GROJ1750_log_observations}).

The next \textit{Chandra} observation was obtained 68 days later and the source was not detected. We obtained an inferred XRT count rate upper limit of <\,7.8$\times$10$^{-5}$\,counts~s$^{-1}$ (see Table\,\ref{tab:P3_GROJ1750_log_observations} for the original \textit{Chandra} count-rate upper limit). After 156 days of our first observation, the source was marginally detected again in the next \textit{Chandra} observation (ObsID 16725; see Fig.\,\ref{fig:P3_GROJ1750_image}, middle panel) with a net count rate of (2.8$\pm$1.2)$\times10^{-4}$\,counts~s$^{-1}$ (8.1$\pm$3.3 net source photons in the 0.5-7\,keV energy range) which resulted in an inferred XRT count rate of $\sim$7.7$\times$10$^{-5}$\,counts~s$^{-1}$.

During our next {\it Chandra} observation (ObsID 16726), one year after the our first observation, the source was detected as well (see Fig.\,\ref{fig:P3_GROJ1750_image}, right panel) with a net count rate of (5.1$\pm$1.5)$\times10^{-4}$\,counts~s$^{-1}$ (14.7$\pm$4.3 net source photons; 0.5-7\,keV) giving an inferred XRT count rate of $\sim$1.5$\times$10$^{-4}$\,counts~s$^{-1}$. Our last \textit{Chandra} observation was obtained approximately two years later but, once again, the source was not detected resulting in an inferred XRT upper limit of <\,5.7$\times$10$^{-5}$\,counts~s$^{-1}$. 
We stacked the 2 observations were no source was detected, but this did not result in a conclusive detection\footnote{We note that only 2.3 net photons were detected at the position of GRO~J1750-27 which does not constitute a significant detection.}. We obtained an inferred XRT upper limit of <\,3.9$\times$10$^{-5}$\,counts~s$^{-1}$.

Before the 2015 outburst, \textit{Chandra} observed GRO~J1750-27 to determine its quiescent luminosity (PI: Wijnands). However, the exposure time was very short ($\sim$5\,ks) and consequently the source was not detected during this observation, with a count-rate upper limit of <\,1.2$\times$10$^{-3}$\,counts~s$^{-1}$ (see Table\,\ref{tab:P3_GROJ1750_log_observations}), resulting in an inferred XRT count-rate upper limit of <\,3.5$\times$10$^{-4}$\,counts~s$^{-1}$. This upper limit is indicated as a dotted brown line in Fig.\,\ref{fig:P3_GROJ1750_BATlightcurve}. However, it is not very constraining since the count rates in the \textit{Chandra} detections were lower than this pre-outburst value (as well as the upper limits obtained from the observations in which we did not detect the source).

\subsection{\textit{Chandra} spectral analysis}\label{subsec:P3_GROJ1750_spectra}

For the three \textit{Chandra} observations during which GRO~J1750-27 was conclusively detected, we obtained the source spectra. The source photons were extracted using a circular region with a radius of $1\,.\!\!^{\prime\prime}5$ centered on the new source position we previously mentioned (see Sect. \ref{subsec:P3_GROJ1750_observations}).  The background photons were extracted using an annulus region (centered on the same position) with an inner and outer radii of $10~~\!\!^{\prime\prime}$ and $20~~\!\!^{\prime\prime}$, respectively (see Fig.\,\ref{fig:P3_GROJ1750_image}).  We used the CIAO tool SPECEXTRACT to obtain the source and background spectra, as well as the response files. We grouped the spectra to 1 count per bin using GRPPHA. The spectra were fitted using XSPEC (v.\,12.9.0)\footnote{\url{https://heasarc.gsfc.nasa.gov/xanadu/xspec/}} in the 0.5-10\,keV energy range using W-statistics (valid for background subtracted spectra). 

We fitted two main basic one-component models to the spectra: an absorbed power-law model (PEGPWRLW) and a blackbody model (BBODYRAD). In addition, we also fitted the spectra using a neutron star atmosphere model applicable for magnetised neutron stars (NSA; see Appendix\,\ref{Appendix:P3_GROJ1750_NSA_model} for the details and results obtained using such a model). This model could also adequately describe  our {\itshape Chandra} spectra, however, in the rest of our paper we mainly use the results obtained when using the power-law and blackbody models in order to be able to compare our source with the other two sources (4U~0115+63 and V~0332+53), for which similar studies have been performed. However, only blackbody and power-law fit results have reported for these two sources 
\citep[see][]{Wijnands2016,Rouco2017a}. Nevertheless, for completeness, we list our NSA model results in Appendix\,\ref{Appendix:P3_GROJ1750_NSA_model}.

For the absorption component, we used TBABS assuming WILM abundances \citep{Wilms2000} and VERN cross-sections \citep{Verner1996}. Since our spectra have very few counts we could not constrain the column density from our spectral fits (see details at the end of this section) and therefore we fixed it to the expected Galactic value in the direction of GRO~J1750-27 (1.03$\times$10$^{22}$\,cm$^{-2}$; \citealt{Kalberla2005}). In the case of the blackbody model, we left the emitting region radius and the temperature as free parameters and determined the unabsorbed 0.5-10\,keV flux by using the convolution model CFLUX. For the power-law model, the energy boundaries were set to 0.5-10\,keV, so that we could directly obtain the unabsorbed flux in that energy range from the model normalization. Due to the low quality of the spectra, both models could fit the data adequately and we could not determine which of the two models is preferable\footnote{We followed the method described by \citet{Tsygankov2017b} to determine if one of the two spectral models used was preferred over the other. In our case, the difference between the values from the W-statistics (C-values) for the power-law and blackbody models was $|\Delta C|$\,<\,2. This difference is below the critical value indicated in \citet[][$|\Delta C|$\,=\,10]{Tsygankov2017b}, therefore we could not statistically prefer one model over the other.}. The results obtained from our spectral fits for both the power-law and the blackbody models are listed in Table\,\ref{tab:P3_GROJ1750_spectral_results}.

\begin{figure*}
	\centering
	\includegraphics[width=2.0\columnwidth]{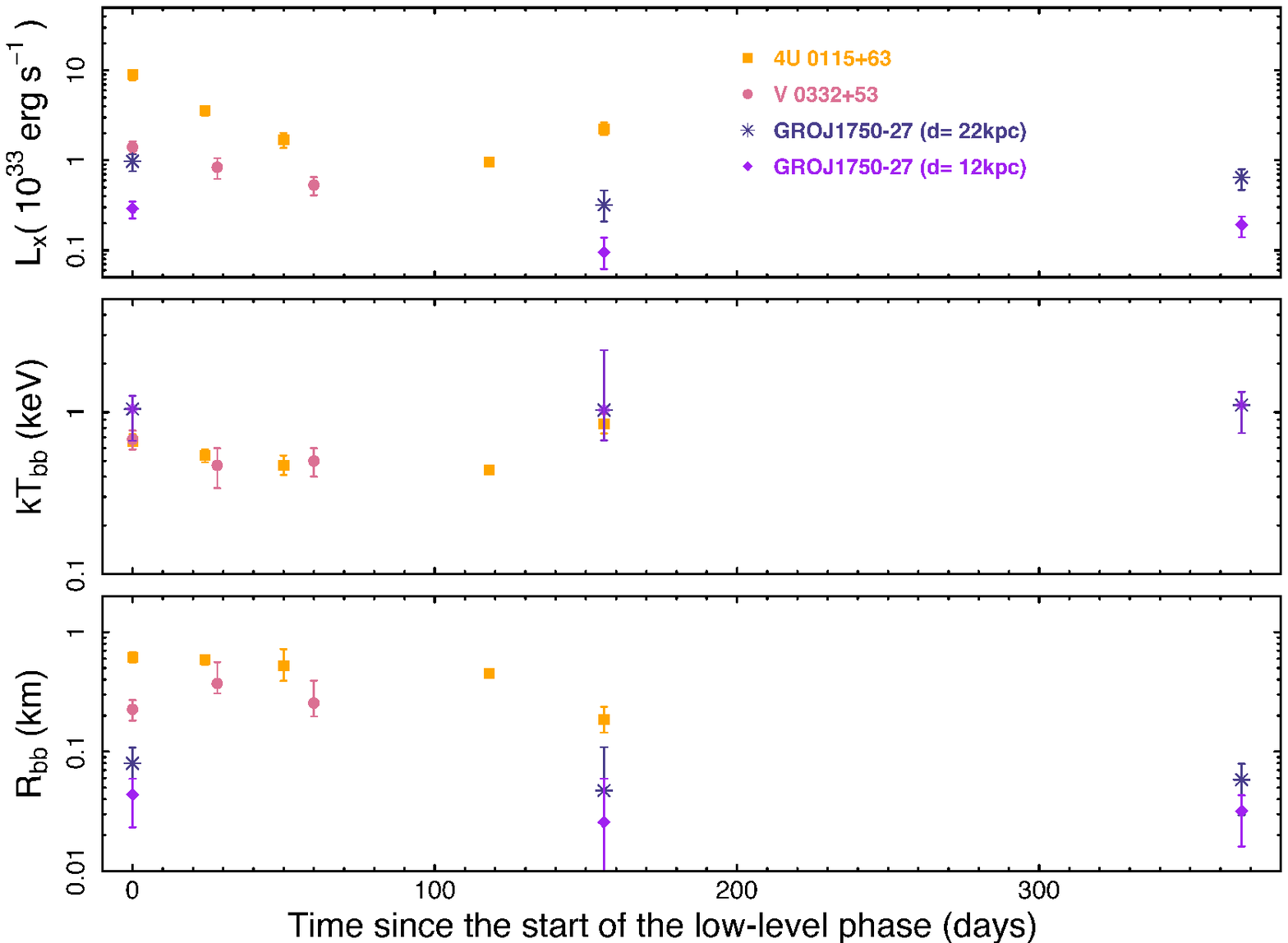}
    \caption{Evolution (using a blackbody model) of the spectral parameters (from the top to the bottom): the X-ray luminosity (for the energy range 0.5-10\,keV), the blackbody temperature, and the associated emission radius. The orange squares are the {\it Swift}/XRT spectral results of 4U~0115+63 published by \citet{Wijnands2016} and \citet{Rouco2017a}. The pink circles correspond to the {\it Swift}/XRT spectral results of V~0332+53 reported by \citet{Wijnands2016}. Luminosities and emission region radii from both sources have been recalculated using their new \textit{Gaia} distances as given in Table\,\ref{tab:P3_GROJ1750_system_parameters}. The dark stars and purple diamonds are our \textit{Chandra} spectral results of GRO~J1750-27 when assuming a source distance of 22\,kpc or 12\,kpc, respectively. Errors are 1$\sigma$. Some points of 4U~0115+63 and V~0332+53 are plotted with symbols that are larger than the corresponding error bars of these points.}
    \label{fig:P3_GROJ1750_bbody}
\end{figure*}

\begin{figure*}
	\centering
	\includegraphics[width=2.0\columnwidth]{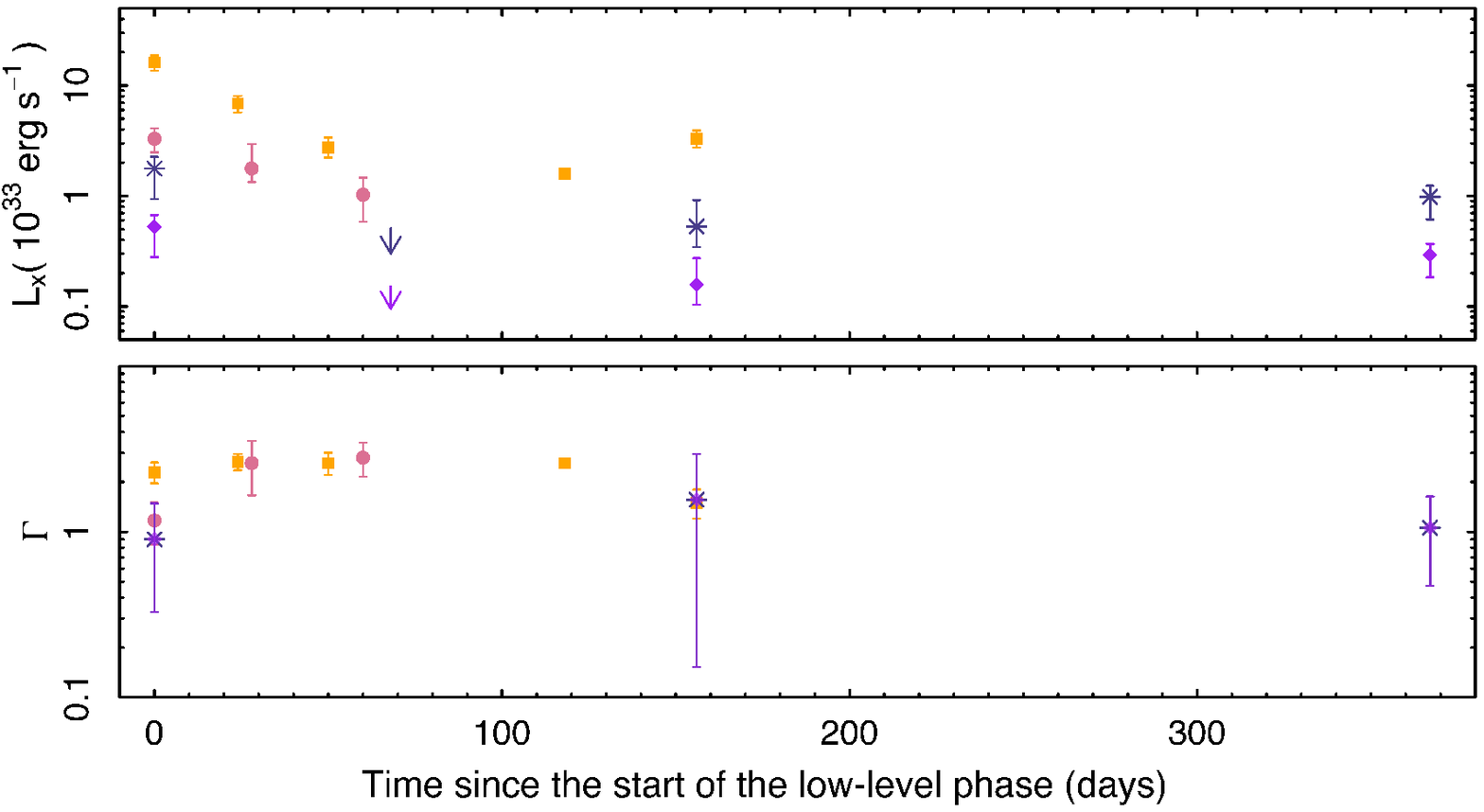}
    \caption{Evolution (using a power-law model) of the spectral parameters: the X-ray luminosity (for the energy range 0.5-10\,keV; top panel) and the photon index (bottom panel). The colours and symbols represent the same sources as in Fig.\,\ref{fig:P3_GROJ1750_bbody}. Errors are 1$\sigma$. Some points of 4U~0115+63 and V~0332+53 are plotted with symbols that are larger than the corresponding error bars of these points. The arrows indicate upper limits on the luminosity and are calculated using the method described in Sect.\,\ref{subsec:P3_GROJ1750_spectra}.}
	\label{fig:P3_GROJ1750_pwl}
\end{figure*}

The source distance is highly uncertain, with estimates ranging from 12\,kpc to 22\,kpc\footnote{\citet{Lutovinov2019} constrained the distance towards the source to a range of 14 to 22\,kpc, very similar to what we assume in our paper.} \citep[see][and references therein]{Shaw2009}. Unfortunately, the source is not detected with \textit{Gaia} so we could not improve on the source distance ourselves (see also Appendix\,\ref{Appendix:P3_GROJ1750_distances}). Therefore, we calculated the luminosity (for both models)\footnote{For the blackbody model, the 0.5-10 keV luminosities are close to the bolometric luminosities, however, for the power-law model the bolometric luminosity could be significantly larger than the 0.5-10 keV luminosity. We do not know the exact spectral shape above 10 keV (in the power-law model) and calculating bolometric luminosities would required additional, very uncertain assumptions about the quiescent source behaviour.  Therefore, we only quote the 0.5-10 keV luminosities. This also allow for direct comparisons with 4U 0115+63 and V 0332+53 for which only the 0.5-10 keV luminosities were given \citep{Wijnands2016, Rouco2017a}.} and the radius of the emission region (for the blackbody model) using both distances. This resulted in X-ray luminosity for the source of L${_X}$$\sim$0.9-3$\times10^{32}$\,erg~s$^{-1}$ or L${_X}$$\sim$0.3-1$\times10^{33}$\,erg~s$^{-1}$ for the blackbody model (see Fig.\,\ref{fig:P3_GROJ1750_bbody}) and L${_X}$$\sim$1.6-5.3$\times10^{32}$\,erg~s$^{-1}$ or L${_X}$$\sim$0.5-1.8$\times10^{33}$\,erg~s$^{-1}$ for the power-law model (see Fig.\,\ref{fig:P3_GROJ1750_pwl}) assuming a distance of 12\,kpc or 22\,kpc, respectively. In the case of the blackbody model, there was no clear evolution in the temperature of the source (kT$_\textnormal{bb}$$\sim$1.1~keV; see Fig.\,\ref{fig:P3_GROJ1750_bbody}); the temperatures measured for the three observations were consistent with each other. Similarly, the radii of the emission regions R$_\textnormal{bb}$$\sim$26-44\,m for 12\,kpc and R$_\textnormal{bb}$$\sim$47-80\,m for 22\,kpc) were consistent within the errors (see Table\,\ref{tab:P3_GROJ1750_spectral_results} and Fig.\,\ref{fig:P3_GROJ1750_bbody}). The inferred radii are much smaller than the NS radius which would suggest that, if the blackbody model is a correct description of the spectra, the emission likely came from a small region on the NS surface, e.g., from hot spots at the magnetic poles (this conclusion also holds if we fit the spectra using a neutron star atmosphere model; see Appendix\,\ref{Appendix:P3_GROJ1750_NSA_model}). In the case of the power-law model, the results from the fit showed that the observed spectra are relatively hard with photon indices ($\Gamma$) of $\sim$0.9-1.6 (see Fig.\,\ref{fig:P3_GROJ1750_pwl} and Table\,\ref{tab:P3_GROJ1750_spectral_results}; as also suggested by the relatively high blackbody temperatures).

For the three \textit{Chandra} observations during which the source was not detected, we converted the obtained count-rate upper limits (see Sect\,\ref{subsec:P3_GROJ1750_observations} and Table\,\ref{tab:P3_GROJ1750_log_observations}) into flux upper limits using the WEBPIMMS tool assuming a power-law model and the spectral parameters obtained from our first \textit{Chandra} detection (observation with ObsID 16723; see Table\,\ref{tab:P3_GROJ1750_spectral_results}). We only used the power-law model because for this model we had to assume only one unknown parameter, i.e., the photon index, to obtain the flux. In the case of the blackbody model, we would have to assume values for two parameters, i.e., the radius of the emission region and its temperature (which are strongly inter-dependent) in order to determine the flux upper limits. Therefore, any upper limit determined using the blackbody model would be more affected by systematic uncertainties than one obtained using the power-law model. After obtaining the flux upper limits, we calculated the luminosity upper limits again assuming 12 and 22\,kpc (see Table\,\ref{tab:P3_GROJ1750_spectral_results} and Fig.\,\ref{fig:P3_GROJ1750_pwl}). The luminosity upper limits during the non-detection observations and the stacked one were significantly below the detection level showing that the source was fainter during these observations and indicating that the system exhibited variability in quiescence.

%
\renewcommand{\arraystretch}{1.5}
\begin{table*}[h]
    \caption{Results of our \textit{Chandra} spectral analysis}
    \centering
    \resizebox{\textwidth}{!}{\begin{tabular}{ccc|ccc|cccc}
        \hline\hline
        \multicolumn{3}{c|}{Detection}    &   \multicolumn{3}{c}{Power-law}   &   \multicolumn{4}{|c}{Blackbody}   \\
        \hline
        ObsID   &   Exposure    &   Distance    &   $\Gamma$    &   F$_\textnormal{X}$  &   L$_\textnormal{X}$  & kT$_\textnormal{bb}$ & R$_\textnormal{bb}$ & F$_\textnormal{X}$ & L$_\textnormal{X}$\\
            &   (ks)    &   (kpc)   &   &   (10$^{-14}$\,erg~cm$^{-2}$~s$^{-1}$) &   (10$^{32}$\,erg~s$^{-1}$)  &   (keV)   &   (10$^{-2}$\,km)  &   (10$^{-14}$\,erg~cm$^{-2}$~s$^{-1}$)  &   (10$^{32}$\,erg~s$^{-1}$)\\
        \hline
        16723   &    26.9   &   12  &   0.90$^{+0.57}_{-0.58}$  &   3.1$^{+1.4}_{-0.9}$ &   5.3$^{+2.5}_{-1.5}$ &   1.05$^{+0.38}_{-0.22}$  &   4.4$^{+2.0}_{-1.6}$ &   1.69$^{+0.38}_{-0.33}$  &   2.91$^{+0.66}_{-0.57}$\\
        "   &   "   &   22  &   "   &   "   &   17.7$^{+8.3}_{-4.9}$    &   "   &   8.0$^{+3.7}_{-2.8}$ &   "   &   9.8$^{+2.2}_{-1.9}$ \\
        16725   &   28.6  &   12  &   1.6$\pm1.4$   &   0.91$^{+0.67}_{-0.32}$  &   1.6$^{+1.2}_{-0.6}$  &   1.0$^{+1.4}_{-0.4}$  &   2.6$^{+3.4}_{-2.6}$ &   0.55$^{+0.25}_{-0.19}$  &   0.95$^{+0.43}_{-0.33}$    \\
		"   &   "   &   22  &   "   &   "   &   5.3$^{+3.9}_{-1.8}$ &   "   &   4.7$^{+6.2}_{-4.7}$ &   "   &   3.2$^{+1.4}_{-1.1}$ \\
        16726   &  " &   12  &   1.06$^{+0.59}_{-0.58}$  &   1.69$^{+0.64}_{-0.44}$  &   2.9$^{+1.1}_{-0.8}$ &   1.11$^{+0.37}_{-0.23}$  &   3.2$^{+1.6}_{-1.1}$ &   1.12$^{+0.31}_{-0.26}$  &   1.92$^{+0.53}_{-0.45}$ \\
        "   &   "   &   22  &   "   &   "   &   9.8$^{+3.7}_{-2.6}$ &   "   & 5.8$^{+2.9}_{-2.1}$ &   "   &   6.5$^{+1.8}_{-1.5}$ \\
    \hline
    \hline 
        \end{tabular}
        }
    \footnotesize   
    \begin{tabular}{ccc|ccc}
        \\
            
        \hline\hline
        \multicolumn{3}{c|}{Upper limit}    &   \multicolumn{3}{c}{Power-law}   \\
        \hline
        ObsID   &   Exposure    &   Distance    &   $\Gamma$    &   F$_\textnormal{X}$  &   L$_\textnormal{X}$  \\
        &   (ks)    &   (kpc)   &   &   (10$^{-14}$\,erg~cm$^{-2}$~s$^{-1}$) &   (10$^{32}$\,erg~s$^{-1}$)  \\
        \hline
        14643   &   4.6 &   12    &   0.90 (fixed)    &   <\,4.1  &   <\,7  \\
        "   &   "   &   22  &   "   &   "   &   <\,23.5 \\
		16724   &   27.3    &   12  &   "   &   <\,0.9  &   <\,1.5  \\
		"   &   "   &   22  &   "   &   "   &   <\,5.1  \\
		16727   &   "   &   12  &   "   &   <\,0.7  &   <\,1.1  \\
		"   &   "   &   22  &   "   &   "   &   <\,3.8  \\	
		167[24,27]  &   54.5  &   12 &   "   &   <\,0.5 &   <\,0.8 \\
		"   &   "   &   22 &   "   &   "   &   <\,2.6 \\
		\hline\hline
    \end{tabular}
    \tablefoot{The N$_\textnormal{H}$ was fixed to 1.03$\times$10$^{22}$\,cm$^{-2}$ (\citealt{Kalberla2005}). All the spectral parameters have been calculated when fitting the spectra in the 0.5-10\,keV energy range. F$_\textnormal{X}$ and L$_\textnormal{X}$ represent the unabsorbed X-ray flux (0.5-10\,keV) and X-ray luminosity (0.5-10\,keV) respectively. The errors are 1$\sigma$. The flux and luminosity upper limits are calculated as mentioned in Sect.\,\ref{subsec:P3_GROJ1750_spectra}.}
    \label{tab:P3_GROJ1750_spectral_results}
\end{table*}
\renewcommand{\arraystretch}{1.0}

%

%
%

%
%

%
\begin{figure*}[h]
	\centering
	\includegraphics[width=2.0\columnwidth]{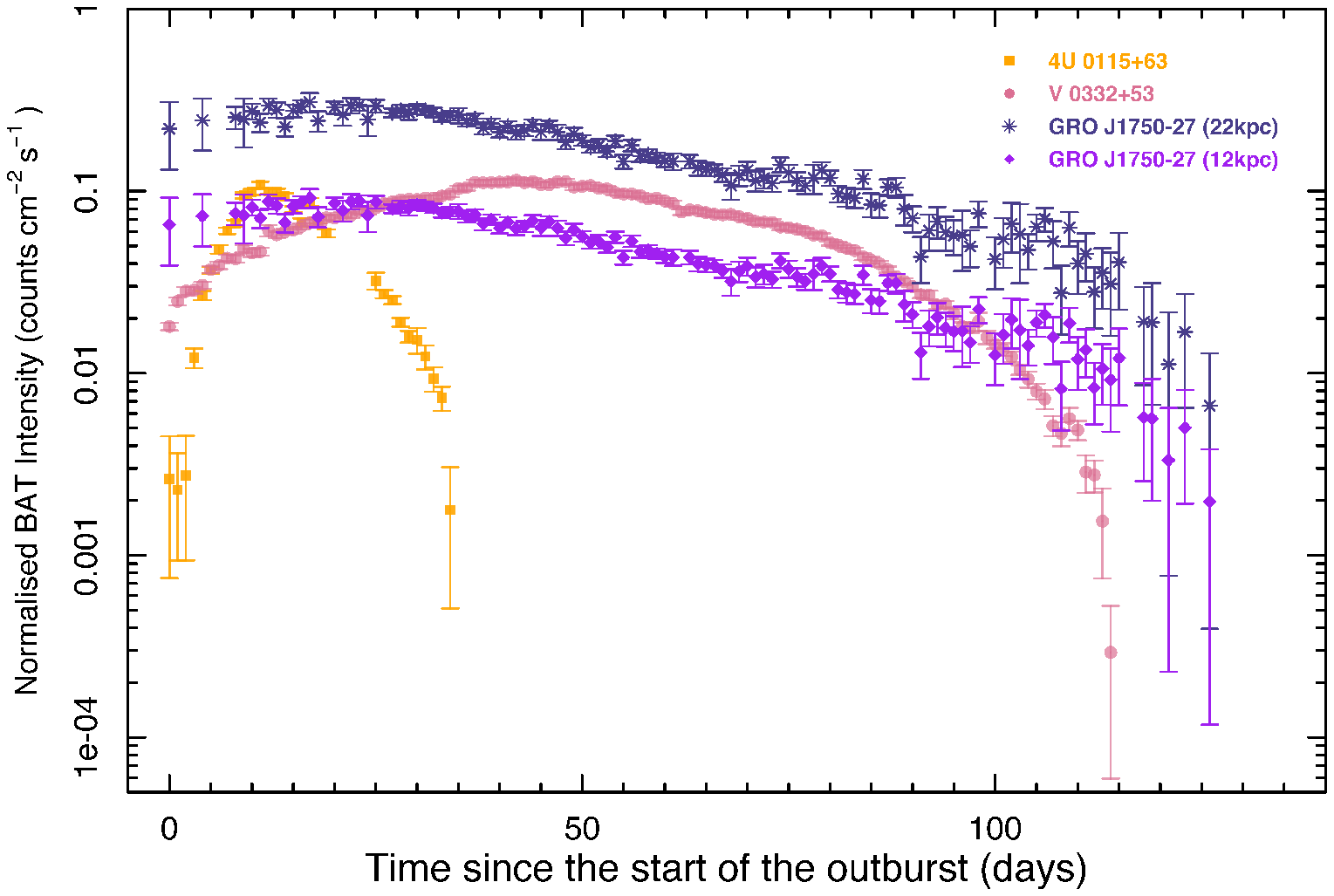}
    \caption{The \textit{Swift}/BAT light curves (in the energy range 15-50\,keV) of the three Be/X-ray transients discussed in the paper. The orange squares correspond to 4U~0115+63, the pink circles to V~0332+53, the dark stars to GRO~J1750-27 assuming that the source is located at a distance of 22\,kpc, and the purple diamonds correspond to GRO J1750-27 for an assumed distance of 12\,kpc. The BAT count rates of the three sources have been normalised to a distance of 7\,kpc using the best estimates for the true source distances (see Table\,\ref{tab:P3_GROJ1750_log_observations}). The start times of the different outbursts are listed in Table \ref{tab:P3_GROJ1750_log_observations}.}
	\label{fig:P3_GROJ1750_fluence}
\end{figure*}
%

Since the N$_{\textnormal{H}}$ is an important parameter during the spectral fitting and nothing is currently known about which exact absorption value for this parameter to be used during the quiescent state of GRO~J1750-27, we tried to constrain N$_{\textnormal{H}}$ from our spectral fits.  For that, we assumed that the N$_{\textnormal{H}}$ has always the same value for all our {\it Chandra} observations. Assuming this, we simultaneously fit the spectra of the three \textit{Chandra} detections with N$_{\textnormal{H}}$ tied between the spectra but it was left free during the fits. In the case of the power-law model, the obtained N$_{\textnormal{H}}$ was $\sim$1.7$\times$10$^{22}$\,cm$^{-2}$ with a (1$\sigma$) confidence interval of [0$-$4.5]$\times$10$^{22}$\,cm$^{-2}$ and for the blackbody model N$_{\textnormal{H}}$  $\sim$1.4$\times$10$^{17}$\,cm$^{-2}$ with a confidence interval of [0$-$1.9]$\times$10$^{22}$\,cm$^{-2}$. Therefore, the N$_{\textnormal{H}}$ is not very well constrained (i.e., for the power-law model) and this is reflected in a significant increase in the errors on the other spectral parameters. However, the obtained fluxes (and thus the inferred luminosities) and their errors are not strongly affected by these uncertainties in the N$_{\textnormal{H}}$. The Galactic N$_{\textnormal{H}}$ towards the source is included within the confidence-interval range obtained when we left the N$_{\textnormal{H}}$ free in the fits, so we decided to fix the N$_{\textnormal{H}}$ to the Galactic value. We note that any different, assumed N$_{\textnormal{H}}$ in the allowed confidence interval will systematically change our spectral parameters slightly (e.g., the blackbody temperature increases by 15\%-20\% while the radius decreases by 20\%-30\% if we assume a N$_{\textnormal{H}}$ close to the maximum allowed value in the blackbody model), but does not alter significantly the fluxes and inferred luminosities so the luminosity difference between GRO~J1750-27 and the other two sources is robust.

\section{Discussion}\label{sec:P3_GROJ1750_discussion}

We present our \textit{Chandra} monitoring campaign of the Be/X-ray transient GRO~J1750-27 after its giant, type-II outburst in 2015. The purpose of our campaign was to determine if the crust of the NS in this system was significantly heated during this outburst and, if so, to follow its crust cooling behaviour. Such cooling of an accretion-heated crusts may have been observed for two other Be/X-ray transients (4U~0115+63 and V~0332+53) after the type-II outbursts they exhibited \citep{Wijnands2016,Rouco2017a}. However, contrary to what was found for these two systems, we do not see any strong evidence of such crust heating and cooling behaviour in GRO~J1750-27 and, consequently, we infer that the NS crust was not significantly heated during the previous outburst (see Section \ref{subsec:P3_GROJ1750_cooling}). We do detect GRO~J1750-27 in three of our five \textit{Chandra} monitoring observations but the spectral parameters (the spectra are relatively hard) as well as the variability seen in quiescence (also taking into account the non-detections) argue that we likely see X-ray emission due to low-level accretion (see Section \ref{subsec:P3_GROJ1750_accretion}). Alternatively, the observed X-ray emission could potentially be originated by the companion star, which, most probably, is an early B-type star (see \citealt{Lutovinov2019}). Such stars are known to be (variable) X-ray emitters as well (e.g., \citealt{Naze2009,Naze2011}). However, as discussed in detail in \citet[][i.e., their Section 4.4]{Tsygankov2017b} the X-ray luminosities of such stars are at most a few times $10^{31}$\,erg~s$^{-1}$. This is significantly below the X-ray luminosities we detect for GRO~J1750-27 and, therefore we suggest that the donor does not, or only very marginally, contribute to the observed X-ray radiation.

\subsection{No heated NS crust in GRO~J1750-27?} \label{subsec:P3_GROJ1750_cooling}

%
\begin{table*}[h]
	\caption{Basic parameters of the systems and their type-II outbursts properties discussed in this paper.}
	\centering
	\resizebox{\textwidth}{!}{\begin{tabular}{cccccccccc}	
		\hline\hline
		\noalign{\smallskip}
		 & Spin & Magnetic Field & Orbital & & & & Start\,--\,End & Outburst & \\
		Name & Period & Strength & Period & Distance & Companion & Eccentricity & Date & Duration & Fluence\\
		 & ($s$) & (10$^{12}$\,G) & (days) & (kpc) & Type & & (MJD) & (days) & (10$^6$\,counts~cm$^{-2}$)\\
		\noalign{\smallskip}
		\hline
		\noalign{\smallskip}
		4U~0115+63 & 3.62$^{a}$ & 1.3$^{b}$ & 24.3$^{c}$ & 7.2$^{d}$ & B0.2Ve$^{e}$ & 0.34$^{e}$ & 57307\,--\,57341 & 34 & $\sim$0.14\\
		V~0332+53 & 4.37$^{f}$ & 3.0$^{g}$ & 33.9$^{h}$ & 5.1$^{d}$ & O8-9Ve$^{i}$ & 0.37$^{h}$ & 57193\,--\,57307 & 114 & $\sim$0.83\\
		GRO~J1750-27 & 4.45$^{j}$ & 2.0\,--\,4.5$^{k, l}$ & 29.8$^{m}$ & 12-22$^{d, k, l}$ & B0-2Ve?$^{l}$ & 0.36$^{m}$ & 57020\,--\,57146 & 126 & $\sim$0.50\,--\,1.67\\
		\noalign{\smallskip}
		\hline\hline
	\end{tabular}
    }
    \tablefoot{The fluence during outburst is given as a relative value (in instrument units) and it is normalised assuming a distance of 7\,kpc for each source.}
    \tablebib{$^{a}$~\citet{Cominsky1978}; $^{b}$~\citet{Raguzova2005}; $^{c}$~\citet{Rappaport1978}; $^{d}$~See Appendix~\ref{Appendix:P3_GROJ1750_distances}; $^{e}$~\citet{Negueruela2001}; $^{f}$~\citet{Stella1985}; $^{g}$~\citet{Makishima1990}; $^{h}$~\citet{Doroshenko2016}; $^{i}$~\citet{Negueruela1999}; $^{j}$~\citet{Bildsten1997}; $^{k}$~\citet{Shaw2009}; $^{l}$~\citet{Lutovinov2019}; $^{m}$~\citet{Scott1997}}.
    \label{tab:P3_GROJ1750_system_parameters}
\end{table*}
%

In the crust heating and cooling scenario, the difference in behaviour of GRO~J1750-27 and the other two systems is unexpected if one looks at their NS parameters (i.e., spin periods and surface magnetic field strengths). For all the systems, these properties are very similar (see Table\,\ref{tab:P3_GROJ1750_system_parameters}) and therefore one would expect, maybe naively, a similar response of the NSs to the accretion of mass. However, it might be that the outbursts of the three sources are significantly different and that might cause the NSs to react differently. As we see in Figure\,\ref{fig:P3_GROJ1750_fluence}, the outburst of GRO~J1750-27 was longer than those of the other two sources and it was, at least, about equally bright (for an assumed distance of 12\,kpc) or even significantly brighter than the outbursts of the other two sources (if GRO\,J1750-27 is located at a large distance of 22\,kpc). Therefore, over the course of the outburst, GRO~J1750-27 seems to have accreted more mass than the other two sources, and thus more energy was liberated in the crust of the NS in GRO~J1750-27 than in that of the other two systems. This makes even more unclear why GRO~J1750-27 did not show any evidence for an accretion-heated NS crust.

To quantify this further, we can compare the fluences of the three outbursts involved. To do this, we need the bolometric luminosities exhibited by the sources during their outbursts. Unfortunately, in the case of Be/X-ray binaries, obtaining the correct F$_\textnormal{bol}$ (and thus the bolometric luminosities) is complicated due to both the wide variety in intrinsic spectral shape between sources and the fast evolution of the absorption column during outburst (e.g., see \citealt{Campana2001a, Reig2013}; see \citealt{Shaw2009} for the spectral evolution of GRO~J1750-27 during its 2008 outburst). These combined effects make it hard to infer the correct spectral shape and therefore, the bolometric luminosities. However, since we are interested in a direct comparison between our three sources, we can get a first approximation of their fluences by comparing their BAT light curves (see Fig.\,\ref{fig:P3_GROJ1750_fluence}; normalized to a distance of 7\,kpc). By integrating these light curves, we obtain the BAT fluences for each outburst (the BAT fluences give the energy output not in physical energy units, but in BAT counts units). The resulting BAT fluences are listed in Table\,\ref{tab:P3_GROJ1750_system_parameters} (we used again two assumed distances for GRO~J1750-27; 12 and 22\,kpc). From these fluences it is clear that if GRO~J1750-27 is located far away, it accreted the largest amount of matter of the three sources. Even if the source is located much closer, it would still have accreted significantly more matter during its outburst than 4U~0115+63 and only $\sim$35$\%$ less than V~0332+53. 

We note that a complication in comparing the different sources with each other is the fact that a significant amount of energy injected in the crust (during outburst) might be released from the rest of the NS surface (and not only from the small emission regions inferred from the blackbody spectral fits) as was shown by \citet{Elshamouty2016}. However, this radiation is unobservable because it likely has a temperature that lies below the {\it Chandra} bandpass. Therefore, it is not possible to obtain a reliable estimate for this potential, additional surface emission from any of our sources. However, if indeed the NS crust in GRO~J1750-27 was heated less than in the other two sources, it remains unclear what causes this difference. Since in all three sources we have NSs with relatively high surface magnetic-field strengths of $\sim$1-4.5$\times$10$^{12}$\,G (see Table\,\ref{tab:P3_GROJ1750_system_parameters}), it might be that the magnetic field inside of the crusts plays an important role. It might be that the configuration and/or the strength of the magnetic field in the NS crust of GRO~J1750-27 is such that it prohibits the crust to show up as significantly heated (see, e.g., the discussions in \citealt{Rouco2017a} and \citealt{Wijnands2017}). 

Alternatively, the NS in GRO~J1750-27 might have accreted significantly less matter over its lifetime than the NSs in the other two systems. As a consequence its crust might not be fully replaced yet with accreted matter (it might have a partially accreted crust; the so-called `hybrid crust'; \citealt{Wijnands2013}), inhibiting some or most of the deep crustal reactions that generate the heating energy (e.g., see also \citealt{Fantina2018}). This would result in a much less heated crust in GRO~J1750-27 than in the other two systems. Finally, it is also possible that the explanation of the differences between the sources lays in, what is called, the `shallow-heating mechanism'. For the low-magnetic field NSs in low-mass X-ray binaries, it has been found that for most of them, the crust cooling curves observed after their outbursts can only be explained if during the accretion in outbursts not only the deep crustal heating reactions occur, but also another heating mechanism is active at shallow depths in the crusts (i.e., $\lesssim$\,150\,m). It has been found that the amount of heating necessary from this mechanism can vary significantly between sources and, even within one source, between different outbursts (see \citealt{Deibel2015}; \citealt{Parikh2017b}; \citealt{Ootes2018}). The physical mechanism behind this shallow heating process is not understood (for a detailed discussion see \citealt{Deibel2015}) and therefore it is quite possible that a similar process might be active during outbursts whe Be/X-ray transients are accreting as well. It might thus be possible that the shallow heating process in GRO~J1750-27 was active at a much lower strength than in the other two sources (or not active at all) and, consequently, resulted in a not or hardly heated crust in GRO~J1750-27.

\subsection{Low-level accretion onto the magnetised NSs}\label{subsec:P3_GROJ1750_accretion}

The variable behaviour seen for GRO~J1750-27 after its outburst cannot easily be explained using the cooling hypothesis. Therefore, it seems prudent to investigate other possibilities for the observed quiescent emission. A possible alternative scenario is one in which the observed quiescent phenomena are caused by residual, low-level accretion onto the NS. Variable levels of the accretion rate could be a natural explanation for the quiescent variability observed in GRO~J1750-27. Evidence for such a low-level accretion is demonstrated by the so-called `mini type-I' outbursts seen in 4U~0115+63 and V~0332+53 \citep{Campana2001a,Wijnands2016}, which were observed on top of the general decay trend of their X-ray luminosities. This slowly decaying behaviour could still be due to the cooling of the NS crust, although a slowly decaying accretion rate cannot be excluded either \citep{Wijnands2016,Rouco2017a}. Unfortunately, how low-level accretion onto a magnetised NS would occur is currently not understood, inhibiting us from making strong conclusive statements. Our three sources are spinning rapidly enough (see Table\,\ref{tab:P3_GROJ1750_system_parameters}) that they are expected to be in the propeller regime at the observed X-ray luminosities (see \citealt{Tsygankov2016} for 4U~0115+63 and V~0332+53; GRO~J1750-27 falls in the propeller-effect area of Fig.\,3 in \citealt{Tsygankov2017a}; see also \citealt{Lutovinov2019}). Therefore, these systems should not exhibit any accretion emission as matter should not be able to reach the NS surface. But clearly matter still reaches the NSs in the propeller regime \citep[as also seen for a few other sources; e.g.][]{Orlandini2004,Mukherjee2005}, although it is unclear whether the mechanism that causes this is the same for these mini-type-I outbursts observed in 4U~0115+63 and V~0332+53, and the emission we see in GRO~J1750-27 in its quiescent state. 

Intriguingly, the second and third \textit{Chandra} detections of GRO~J1750-27 (during ObsIDs 16725 and 16726) occurred close to periastron passage (see Table\,\ref{tab:P3_GROJ1750_log_observations}) similar to the mini type-I outbursts in the other two sources\footnote{The \textit{Chandra} observation 16724 was also obtained close to periastron (see Table\,\ref{tab:P3_GROJ1750_log_observations}) but the source was not detected. This indicates that if the emission mechanism is linked to the periastron passage, it is not always active. This is similar to what has been found for the mini type-I outbursts in 4U~0115+63 and V~0332+53 which are not always present at periastron passages \citep[][]{Wijnands2016,Rouco2017a}.}. This might indicate a possible link between the emission mechanisms in the different sources, although the peak luminosities at periastron vary widely between sources: L$_\textnormal{X}$$\sim$10$^{34-35}$~erg~s$^{-1}$ for 4U~0115+63 and V~0332+53 \citep{Campana2001b,Wijnands2016} versus L$_\textnormal{X}$$\sim$10$^{32}$~erg~s$^{-1}$ for GRO~J1750-27. The large range in luminosity might be difficult to explain in any model assuming that the underlying physical mechanism is the same in all sources. In addition, the first \textit{Chandra} detection of GRO~J1750-27 occurred far from periastron (see Table\,\ref{tab:P3_GROJ1750_log_observations}). This might be explained by assuming that this {\it Chandra} observation was performed during the accretion tail from the final stage of the giant outburst (see also \citealt{Lutovinov2019}). Or that the accretion mechanism at work in this source is different for the first observation compared to that during the other two. In this case, it might be possible that two mechanisms are at work in GRO~J1750-27, one related to the mini type-I outburst phenomenon and one that causes accretion when the source is far away from periastron. Clearly, more studies, both observational as well as theoretical, are needed to improve our understanding of low-level accretion onto magnetised NSs.

We note that if our first {\itshape Chandra} observation was indeed obtained during the tail of the type-II outburst, the source might not have been fully in "true" quiescence yet\footnote{Even if the first {\itshape Chandra} observation was performed while the type-II outburst was not fully over yet, the question of how accretion onto a magnetised neutron star occurs at the very low observed accretion rate (as inferred from the very low observed X-ray luminosity during this observation) still remains.}. If this observation is ignored in the light curves, the detected variability during the other four {\itshape Chandra} observations is less significant (i.e., in the light that we only have a handful of photons detected). In order to check whether the source was indeed variable during these four {\itshape Chandra} observations, we fitted the observed net count rates (we used 1$\sigma$ errors) with a constant model. This model could not fit the data well with a $\chi^2$ value of 12.05 for 3 degrees of freedom (leading to a p-value of 0.0072; this p-value decreases to $2.8\times10^{-5}$ when also including our first {\itshape Chandra} point). This suggest that indeed the source was variable during our observations. However, even in the unlikely case that the source was not variable during the last four {\itshape Chandra} observations, our conclusion still holds that the neutron-star crust in GRO~J1750$-$27 was significantly less heated (because of the very low quiescent luminosities observed from our target) during its preceding type-II outburst than the crusts in the other two sources discussed in our paper. In addition, the hard quiescent spectra we observed for our source with respect to the spectra observed for the other two sources (Figures \ref{fig:P3_GROJ1750_bbody} and \ref{fig:P3_GROJ1750_pwl}), also indicate that the emission we detected is indeed due to low-level accretion of matter on the neutron star.

\begin{acknowledgements}\label{sec:P3_GROJ1750_acknowledgements} ARE, LSO, and RW were supported by a NWO Top Grant, module 1 awarded to RW. Support for this work was provided by the National Aeronautics and Space Administration through Chandra Award Number GO5-16040X issued by the Chandra X-ray Observatory Center, which is operated by the Smithsonian Astrophysical Observatory for and on behalf of the National Aeronautics Space Administration under contract NAS8-03060. This work has benefited from support by the National Science Foundation under Grant No. PHY-1430152 (JINA Center for the Evolution of the Elements). Data from the European Space Agency (ESA) mission \textit{Gaia} (\url{https://www.cosmos.esa.int/gaia}) have been used in this project and processed by the \textit{Gaia} Data Processing and Analysis Consortium (DPAC, \url{https://www.cosmos.esa.int/web/gaia/dpac/consortium}). Funding for the DPAC has been provided by national institutions, in particular the institutions participating in the \textit{Gaia} Multilateral Agreement.\\
\end{acknowledgements}

%

\bibliographystyle{aa}
\bibliography{references}\label{P3_GROJ1750_references}

%
\newpage
\begin{appendix}
\onecolumn

\section{Gaia distances}\label{Appendix:P3_GROJ1750_distances}

\textit{Gaia} did not detect the optical counterpart of GRO~J1750-27 so we could not improve on the distance estimate for this source. The distances used in this paper for the other two Be/X-ray transients, 4U~0115+63 and V~0332+53, have been obtained from the second \textit{Gaia} data release (GDR2; \citealt{Gaia2016}; \citealt{Gaia2018}). We obtained the parallaxes of the sources from the \textit{Gaia} archive\footnote{\url{http://gea.esac.esa.int/archive/}} and used the code developed by \cite{Bailer-Jones2018}\footnote{\url{https://github.com/ehalley/Gaia-DR2-distances}} to determine the best distance estimates. This code computes the distance using a prior that varies depending on the Galactic longitude and latitude according to a three-dimensional model of the Galaxy \citep{Rybizki2018}. Due to the nonlinear transformation, the confidence intervals on the distances that we obtain are asymmetric. Table\,\ref{tab:P3_GROJ1750_gaia_parameters} shows the input parameters for the code and the estimated distances. The parallaxes were corrected from the zeropoint offset in the catalogue ($+0.029$~mas) as determined from \textit{Gaia} observations of quasars by \cite{Lindegren2018}.

%
\begin{table*}[h]
\caption{Main parameters used for obtaining \textit{Gaia} distances}
	\centering
	\resizebox{\textwidth}{!}{\begin{tabular}{ccccccc}
    	\hline\hline
		\noalign{\smallskip}
		Source & \textit{Gaia} ID & $l$~(J2000) & $b$~(J2000) & Parallax & Estimated Distance & Range Distances\\
		  & & ($^{\circ}$) & ($^{\circ}$) & (10$^{-2}$\,mas) & (kpc) & (kpc) \\
        \noalign{\smallskip}
		\hline
		\noalign{\smallskip}
        4U~0115+63 & 524677469790488960 & 125.924 & 1.028 & 9.1$\pm$2.7 & 7.2 & [6.1-8.7]\\
        V~0332+53 & 444752973131169664 & 146.052 & -2.194 & 14.3$\pm$3.6 & 5.1 & [4.4-6.2]\\
        \noalign{\smallskip}
		\hline
    \end{tabular}
    }
    \tablefoot{From left to right: source name, \textit{Gaia} ID, Galactic longitude and latitude, parallax, the estimate of distance, and the range with the lower and upper bounds at the 68$\%$ asymmetric confidence level interval for the distance following \cite{Bailer-Jones2018}.}
    \label{tab:P3_GROJ1750_gaia_parameters}
\end{table*}
%

%
\begin{figure}[h]
    \centering
    \begin{subfigure}[b]{0.45\textwidth}
        \includegraphics[width=\textwidth]{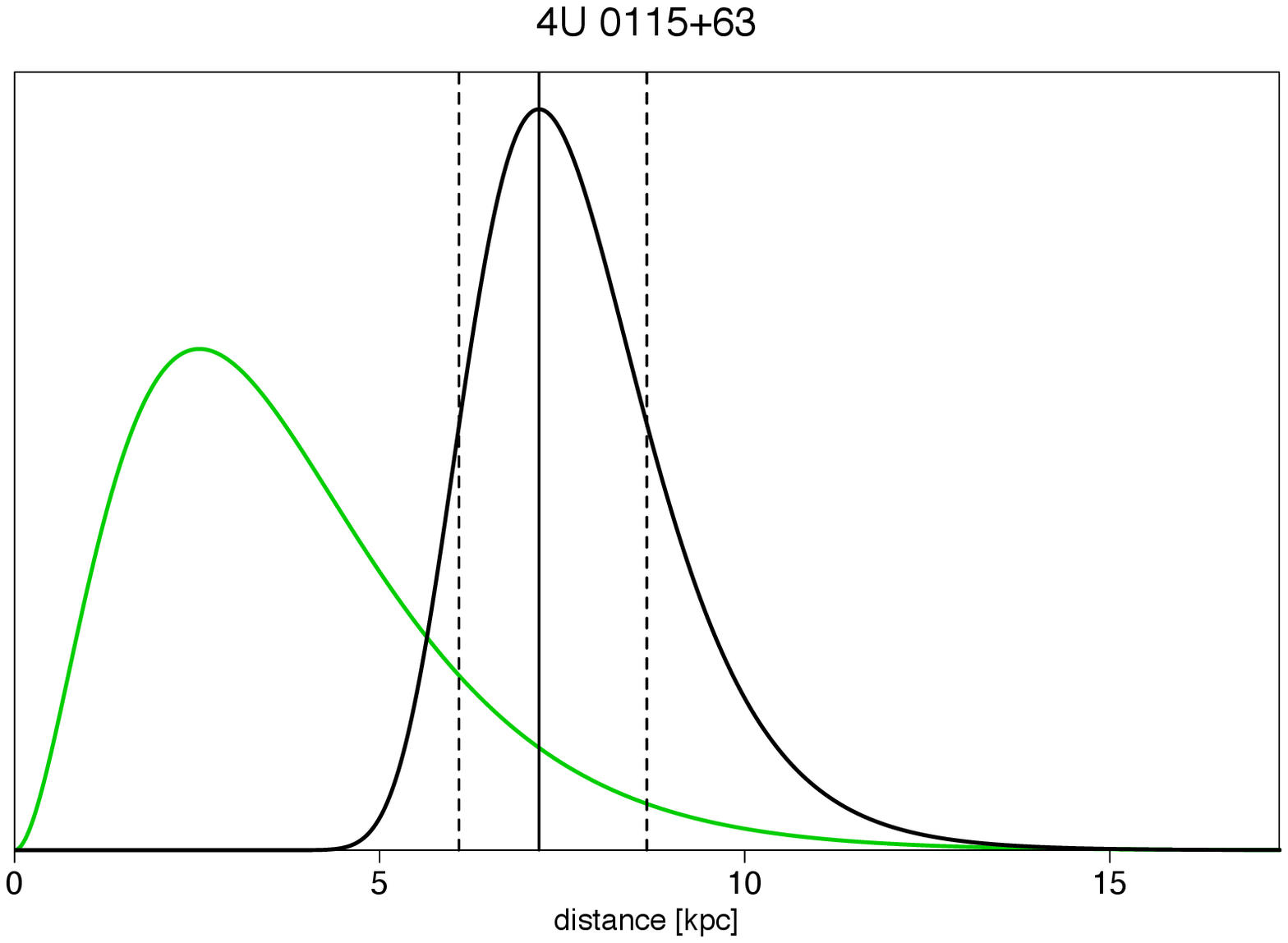}
        \label{fig:4U0115_Gaia_Distance}
    \end{subfigure}
    \begin{subfigure}[b]{0.45\textwidth}
        \includegraphics[width=\textwidth]{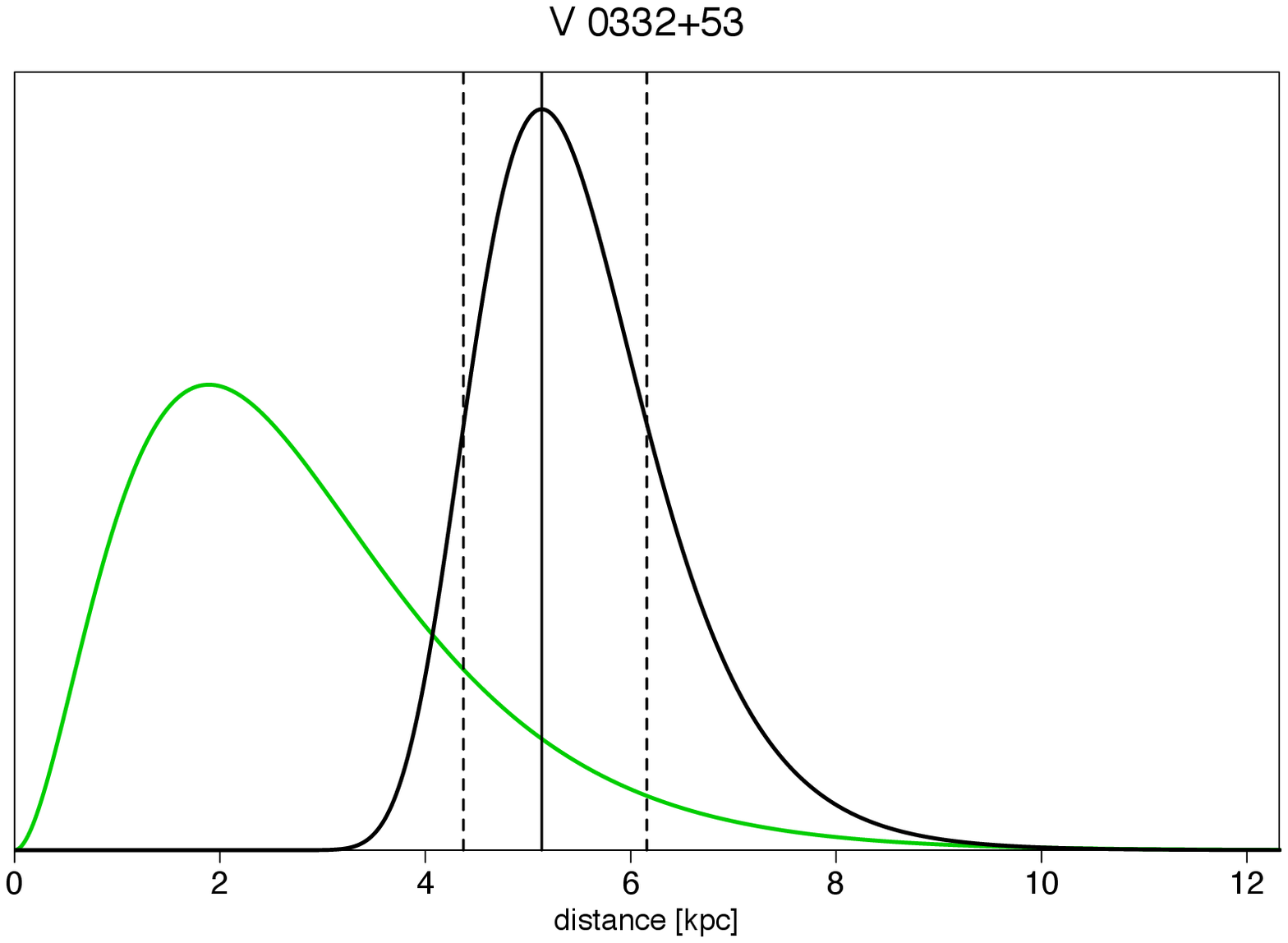}
        \label{fig:V0332_Gaia_Distance}
    \end{subfigure}
    \caption{Plots obtained using the code \citep{Bailer-Jones2018} to determine the best \textit{Gaia} distance estimates for 4U~0115+63 (left panel) and V~0332+53 (right panel). The prior is represented by the green line and the posterior is plotted in black. The black vertical line corresponds to the distance estimator, and the dashed vertical lines are the lower and upper bounds of the 68$\%$ confidence interval.}
    \label{fig:P3_gaiadistance}
\end{figure}
%

\newpage

\section{Spectral fit results using the neutron star atmosphere model (NSA)}\label{Appendix:P3_GROJ1750_NSA_model}

We also fitted our \textit{Chandra} spectra using the neutron star atmosphere (NSA) assuming a magnetised neutron star (\citealt{Pavlov1995}). We used the same spectral analysis set-up as the one described in Sect.\,\ref{subsec:P3_GROJ1750_spectra}, but in addition we fixed the neutron-star mass (1.4\,M$_{\odot}$), radius (11\,km), and magnetic field strength (we used 2$\times$10$^{12}$\,G; the results were very similar if we used 4.5$\times$10$^{12}$\,G so we only report the first set of fit results) values. We left the normalization of the model to vary freely in order to calculate the size of the emission region for both assumed distances (12\,kpc and 22\,kpc). As for the blackbody model, the fluxes were obtained using the CFLUX command. The results of the fit are shown in Table\,\ref{tab:P3_GROJ1750_nsa_parameters}. Both the (unredshifted) effective temperature and the size of the emission region are fully consistent between the three \textit{Chandra} observations, with the fluxes varying between these observations. We note that these inferred sizes for the emission region are much smaller than the NS radius indicative of the existence of hot spots as also was deduced from the blackbody fits (which produced slightly larger emission region sizes; see Table\,\ref{tab:P3_GROJ1750_spectral_results}). The fluxes obtained using the neutron star atmosphere model are fully consistent with those obtained using the blackbody model.

\renewcommand{\arraystretch}{1.5}
\begin{table*}[h]
    \caption{Results of the neutron star atmosphere model fitting}
    \centering
    \begin{tabular}{ccc|cccc}
        \hline\hline
        \multicolumn{3}{c|}{Detection}    &   \multicolumn{4}{c}{B$ = $2$\times$10$^{12}$\,G}  \\
        \hline
        ObsID   &   Exposure    &   Distance    &   LogT$_\textnormal{eff}$ &   R$_\textnormal{emission}$ &   F$_\textnormal{X}$  &   L$_\textnormal{X}$  \\
            &   (ks)    &   (kpc)   &   (10$^{-2}$\,K) &   (10$^{-2}$\,km)    &   (10$^{-14}$\,erg~cm$^{-2}$~s$^{-1}$) &   (10$^{32}$\,erg~s$^{-1}$)  \\
        \hline
        16723   &   26.9    &   12  &   711.2$^{+5.1}_{-5.5}$  &  2.72$^{+1.31}_{-0.87}$ &    1.63$^{+0.37}_{-0.32}$ &   2.81$^{+0.63}_{-0.55}$ \\
        "   &   "   &   22  &   "   &   5.0$^{+2.4}_{-1.6}$  &   "   &   9.4$^{+2.1}_{-1.9}$    \\
        16725   &   28.6    &   12  &   712$^{+10}_{-12}$  &  1.5$^{+2.2}_{-1.5}$ &    0.54$^{+0.25}_{-0.19}$ &   0.94$^{+0.42}_{-0.33}$ \\
        "   &   "   &   22  &   "   &   2.7$^{+4.0}_{-2.7}$  &   "   &   3.2$^{+1.4}_{-1.1}$    \\
        16726   &   "    &   12  &   712.5$^{+5.9}_{-5.5}$  &  1.94$^{+1.02}_{-0.78}$ &    1.06$^{+0.33}_{-0.28}$ &   1.83$^{+0.57}_{-0.48}$ \\
        "   &   "   &   22  &   "   &   3.6$^{+1.9}_{-1.4}$  &   "   &   6.1$^{+1.9}_{-1.6}$    \\ 
        \hline\hline
        \end{tabular}
    \tablefoot{The value of the absorption column was fixed to 1.03$\times$10$^{22}$\,cm$^{-2}$ (\citealt{Kalberla2005}). The spectral parameters, fluxes and luminosities have been calculated in the 0.5-10\,keV energy range. The errors are 1$\sigma$.}
    \label{tab:P3_GROJ1750_nsa_parameters}
\end{table*}
\renewcommand{\arraystretch}{1.0}

\end{appendix}

\end{document}